\documentstyle[12pt]{article}
\begin{document}

\begin{center}
{\noindent{\bf{Chiral Symmetry Breaking and Dual Gluon Mass in the Confining Region of QCD}}}

\vspace{0.5cm}

Alok Kumar{\footnote{e-mail address: alok@imsc.res.in}}  and R.Parthasarathy{\footnote{e-mail
address:sarathy@imsc.res.in}} \\
The Institute of Mathematical Sciences \\
C.P.T.Campus, Taramani Post \\
Chennai 600 113 \\
India. \\
\end{center}

\vspace{3.5cm}  

{\noindent{\it{Abstract}}}

\vspace{0.5cm}

The Dual Meissner Effect description of QCD in the confining region provides $\frac{1}{q^4}$ behaviour
for the gluon propagator and involves the dual gluon mass $m$ as a parameter. This is used in the
Schwinger-Dyson equation for the quarks in the infrared region to exhibit chiral symmetry breaking for
light quarks. Using the light quark condensate as input, the dual gluon mass is determined and its
importance in showing the asymptotic free behaviour of the extrinsic curvature coupling in the rigid
QCD string is discussed.
\newpage 

\vspace{2.5cm}   

Confinement of quarks and gluons and chiral symmetry breaking are the two important
non-perturbative (infra-red) features of QCD and so any model of confinement must
necessarily produce chiral symmetry breaking nonperturbatively. From the
phenemenological quark-antiquark confining linear potential, most suggestions of
confinement imply $\frac{1}{q^4}$ behaviour for the gluon propagator in the infrared
region. Such a linear potential violates the cluster property in a local quantum field
theory [1]. However, Strocchi [2] demonstrated that the proof of the cluster property
for QED {\it{actually fails for QCD upon realizing that the Yang-Mills field strength
$F^a_{\mu\nu}$ in QCD are not observables}} in contrast to the observability of the
electric and magnetic fields in QED. One of the authors (R.P) motivated by the
suggestion of 'tHooft [3] and Mandelstam [4] that monopoles must play a crucial role
in the infrared regime of QCD and that of Nambu [5] that dual Meissner effect as a
mechanism of confinement, described the infrared regime of $SU(3) QCD$ by proposing
magnetic symmetry condition to select the confining configurations of $A^a_{\mu}$ [6].
By exactly solving this condition, appropriate $SU(3)$ gauge fields relevant to
describe the infrared regime were obtained. A dual QCD action had been derived and the
quantum one-loop corrections generated mass for the dual gluons by 'gauge mixing
mechanism' [7]. The effective action then exhibited condensation of monopoles as in the
London theory of Meissner effect and the dual monopole corresponded to the Abrikosov
flux. Subsequently the
author (R.P) derived a string representation of the above dual action [8] and this
was that of a {\it{rigid string}} involving besides the Nambu-Goto term, an extrinsic
curvature dependent term. A string with extrinsic curvature has been proposed earlier
by Polyakov [9] and Kleinert [10] to describe QCD and has been studied in detail using
generalized Gauss Map [11]. The extrinsic curvature term has gained importance in view of the result
of Kleinert [12] who showed that it is generated {\it{dynamically}} by the fluctuations of the string
world-sheet. Subsequently, Kleinert [13] presented a continuum field theory of compact $U(1)$ gauge
theory which admits monopoles. In contrast to the dual Higgs field description of quark confinement,
he [13] used the original formulation in terms of the compact $U(1)$ gauge field and dualized the
action to describe dual Meissner effect. The dual gauge fields acquire mass and this gives rise to
linear confining potential. Later,    
Polyakov considered the compact $U(1)$ gauge theory
 and obtained rigid string action in the large area limit of
the Wilson loop. So it is natural that the dual Meissner effect description [6] gives
the rigid string action [8]. However in [6] and [8], the dual gluon mass was not determined.  

\vspace{0.5cm}

It is the purpose of this paper first to demonstrate chiral symmetry breaking in QCD
using the confining configurations of [6] that gave the dual Meissner effect and then
use these results to determine the dual gluon mass $m$. The dual gluon mass is an important parameter
in the dual superconductor picture of confining region of QCD and its determination will be very
useful in its own right besides giving the relative strength of the extrinsic curvature coupling in
the QCD string description.   
 The massive dual
gluon propagator (Eqn.15 of Ref.6) in Euclidean version has been found to be
$\frac{{\delta}_{\mu\nu}}{q^2-m^2}$ and the gluon propagator, using the analysis of
Nair and Rosenzweig [15], is $-\frac{m^2}{q^4}{\delta}_{\mu\nu}$ in Feynman gauge. As
the above dual gluon propagator has been obtained in the confining region of QCD, this
gluon propagator is valid only in the infrared region. We will examine the chiral
symmetry breaking using the above propagator for the gluons and Schwinger-Dyson
equation for the quark propagator in the rainbow approximation. The input to determine
the dual gluon mass $m$ will be the light quark condensate. 

\vspace{0.5cm}

Schwinger-Dyson equation provides a nonperturbative analysis of the quark propagator. A
complete solution of this equation is extremely difficult and various approximations
have been used. For a review, see Roberts and Schmidt [16]. The Schwinger-Dyson
equation for quark propagator is
\begin{eqnarray}
S^{-1}(p)&=&Z_2(i\gamma \cdot p +m_b)+Z_1\int \frac{d^4q}{(2\pi)^4}g^2{\gamma}^{\mu}
\frac{{\lambda}^a}{2}S(q){\Gamma}^{\nu b}D^{ab}_{\mu\nu}(p-q),
\end{eqnarray}
where $S(p)$ is the full quark propagator, $D^{ab}_{\mu\nu}$ is the dressed gluon
propagator, ${\Gamma}^{\nu b}$ is the dressed quark-gluon vertex, $m_b$ is the bare
mass of the quark (which will be set equal to zero hereafter), $Z_1$ is the quark-gluon
vertex renormalization and $Z_2$ is the quark wave function renormalization.
In applying (1) to QCD in the infrared region, we invoke the following: (1) In the
infrared region, the QCD coupling $g^2$ will be taken to be a constant, following
Gribov [17]. This kind of 'freezing' of $g^2$ has been pointed out earlier [18,19] and
discussed by Aguilar, Mihara and Natale [20]. Recently, Alkofer and Fischer [21]
observed that the nonperturbative strong running coupling resulting from gluon and
ghost propagators possesses an infrared fixed point. Although the origin of the
infrared fixed point is not known yet, we will treat $g^2$ constant in the infrared
region of QCD. (2) We will take $m_b=0$ so that the $p^2$-dependent mass term in $S(p)$
at $p^2=0$, if not zero, will ensure chiral symmetry breaking. (3) We will use rainbow
approximation and take ${\Gamma}^{\nu b}={\gamma}^{\nu} \frac{{\lambda}^b}{2}$ and
$Z_1=Z_2=1$. Then writing $D^{ab}_{\mu\nu}(p-q)={\delta}^{ab}D_{\mu\nu}(p-q)$, (1)
becomes
\begin{eqnarray}
S^{-1}(p)&=&i\gamma \cdot p +g^2\int \frac{d^4q}{(2\pi)^4}
{\gamma}^{\mu}S(p-q){\gamma}^{\nu}D_{\mu\nu}(q).
\end{eqnarray}
The standard procedure is to assume a solution of (2) in terms of functions $A(p^2)$
and $M(p^2)$ so as to have
\begin{eqnarray}
S(p)\ =\ \frac{1}{i\gamma \cdot p A(p^2)+M(p^2)}&=&i\gamma \cdot p C(p^2)+B(p^2),
\end{eqnarray}
where
\begin{eqnarray}
C(p^2)&=& \frac{-A(p^2)}{p^2A^2(p^2)+M^2(p^2)}, \nonumber \\
B(p^2)&=&\frac{M(p^2)}{p^2A^2(p^2)+M^2(p^2)}, \ \ \ and \ so\ \nonumber \\
p^2C^2(p^2)+B^2(p^2)&=&{\{ p^2A^2(p^2)+M^2(p^2)\}}^{-1}.
\end{eqnarray}
Using (3) in (2) and taking the trace over the gamma matrices, we obtain the following
integral equations for $C(p^2)$ and $B(p^2)$.
\begin{eqnarray}
\frac{B(p^2)}{p^2C^2(p^2)+B^2(p^2)}&=&g^2\int \frac{d^4q}{(2\pi)^4}D^{\mu}_{\mu}(q)
B((p-q)^2), \nonumber \\
\frac{-p^2C(p^2)}{p^2C^2(p^2)+B^2(p^2)}&=&p^2+g^2\int \frac{d^4q}{(2\pi)^4}
C((p-q)^2) \nonumber \\ 
& &\{2p^{\mu}(p-q)^{\nu}-p\cdot (p-q){\delta}^{\mu\nu}\}D_{\mu\nu}(q).
\end{eqnarray} 
The considerations in [6] led to $\frac{1}{q^4}$ behaviour for the gluon propagator in
the infrared region of QCD and we take $D_{\mu\nu}(q)$ in the Feynman gauge as 
\begin{eqnarray}
D_{\mu\nu}^{ab}(q)&=&-\frac{m^2}{q^4}{\delta}_{\mu\nu}{\delta}^{ab}.
\end{eqnarray}
Then the integral equation for $B(p^2)$ becomes
\begin{eqnarray}
\frac{B(p^2)}{p^2C^2(p^2)+B^2(p^2)}&=&-4m^2g^2\int
\frac{d^4q}{(2\pi)^4}\frac{B((p-q)^2)}{q^4}.
\end{eqnarray} 
After a change of variable as $k^{\mu}=(p-q)^{\mu}$, the right side of (7) becomes 
\begin{eqnarray}
&-4m^2g^2\int \frac{d^4k}{(2\pi)^4}\frac{B(k^2)}{((p-k)^2)^2}& \nonumber 
\end{eqnarray}
and carrying out the angular integration, we have this as 
\begin{eqnarray}
&-8{\pi}^2m^2g^2\int \frac{k^3dk}{(2\pi)^4}\frac{B(k^2)}{(p^2-k^2)^2}& \nonumber
\end{eqnarray}     

\vspace{0.5cm}

We shall now consider $B(k^2)$ as an analytic function of $k^2$ and use Cauchy-Riemann
representation
\begin{eqnarray}
B(k^2)&=&\frac{1}{2\pi i}\int _{-\infty}^{\infty} d\alpha
\frac{\bar{B}(\alpha)}{k^2-\alpha}.
\end{eqnarray}
Such a representation is justifiable on the ground that quark fields are not
observables, the observability criteria being taken as the vanishing of the commutator
of the BRST charge $Q_B$ and the corresponding field [22], the physical subspace being
annihilated by $Q_B$. As the commutator of $Q_B$ with quark field is not vanishing,
quarks are not observables. This ensures that the quark fields are not asymptotic
states. Further, the above representation has been used recently by Anishetty and
Kudtarkar [23] in a theory of QCD with string tension. Then, we have
\begin{eqnarray}
\frac{B(p^2)}{p^2C^2(p^2)+B^2(p^2)}&=&-\frac{8{\pi}^2m^2g^2}{2\pi i}\int
\frac{k^3dk}{(2\pi)^4} \nonumber \\
& & \int_{-\infty}^{\infty}d\alpha \frac{\bar{B}(\alpha)}{(k^2-\alpha)(p^2-k^2)^2}.
\end{eqnarray}
Using Feynman parameterization to write
\begin{eqnarray}
\frac{1}{(k^-p^2)^2(k^2-\alpha)}&=&\int_{0}^{1}dx \frac{2x}{[k^2-xp^2-(1-x)\alpha]^3},
\nonumber 
\end{eqnarray}
and carrying out the $dk$ integration, the right side of (9) becomes
\begin{eqnarray} 
& \frac{16{\pi}^2m^2g^2}{4(2\pi i)(2\pi)^4}\int_{-\infty}^{\infty}d\alpha
\bar{B}(\alpha)\int_{0}^{1}\frac{xdx}{xp^2+(1-x)\alpha}.& \nonumber
\end{eqnarray}
The $x$-integration is performed to give the right side of (9) as 
\begin{eqnarray}
&\frac{m^2g^2}{4{\pi}^2}B(p^2)-\frac{m^2g^2}{4{\pi}^2(2\pi i)}\int_{-\infty}^{\infty}
d\alpha \frac{\bar{B}(\alpha)}{(p^2-\alpha)^2}\ell og(\frac{p^2}{\alpha}),& \nonumber 
\end{eqnarray}
where we have used (8) to write the first term. We assume that $\bar{B}(\alpha)$ is a
smooth function of $\alpha$. Then $\alpha=p^2$ is a double pole and $\alpha=0$
singularity in the logarithm will not contribute. Evaluating the double pole
contribution at $\alpha=p^2$, we find
\begin{eqnarray}
\frac{1}{p^2C^2(p^2)+B^2(p^2)}&=& \frac{m^2g^2}{2{\pi}^2},
\end{eqnarray}
and therefore from (4)
\begin{eqnarray}
p^2A^2(p^2)+M^2(p^2)&=&\frac{m^2g^2}{2{\pi}^2}.
\end{eqnarray}    

\vspace{0.5cm}

Thus the unknown functions $A(p^2)$ and $M(p^2)$ in the solution of the Schwinger-Dyson
equation are constrained by (11). This relation obtained in the rainbow approximation
using $\frac{1}{q^4}$ propagator for gluons, as follows from [6], in the Feynman gauge
has not been reported in the literature to the best of our knowledge and so constitute
a new relation. The validity of this relation is only in the infrared region of QCD
described by the confining configurations of Ref.6. It is interesting however to record
that the above relation (10,11) also follows from the $\delta$-function behaviour of
the confined dressed gluon propagator in the non-perturbative region. Such a propagator
has been suggested by Munczek and Nemirovsky [24] as 
\begin{eqnarray}
D^{ab}_{\mu\nu}(q)&=&{\delta}^{ab}({\delta}_{\mu\nu}-\frac{q_{\mu}q_{\nu}}{q^2})(2{\pi})^4{\cal{G}}^2 {\delta}^4(q),
\end{eqnarray}
where ${\cal{G}}$ is a dimensionful constant. This propagator (12) has been extensively
used by Roberts and his co-workers [25] in their study of Schwinger-Dyson equation.
Using this propagator in the first equation of (5), we find
\begin{eqnarray}
p^2C^2(p^2)+B^2(p^2)&=&\frac{1}{3g^2{\cal{G}}^2}\ =\ a\ constant,
\end{eqnarray}
which is same as (10) and so (11) with $m^2=6{\pi}^2{\cal{G}}^2$. This comparison, of
two different gluon propagators giving the same relation, shows the importance of
(10,11) in the confining region of QCD. We wish to state that while the propagator (6)
has a field theoretic derivation in [6] and gives rise to linear potential, the
propagator (12) seems to be an ansatz and will not give rise to linear potential.       

\vspace{0.5cm}

In a similar manner, we find $C(p^2)$ in (5) becomes a constant and this is also the
result of using (13). The constancy of $C(p^2)$ is valid only in the infrared region.
This in turn gives $A(p^2)$ is $\frac{2}{3}$ a dimensionless constant. Thus (11)
becomes 
\begin{eqnarray}
\frac{4p^2}{9}+M^2(p^2)&=& \frac{m^2g^2}{2{\pi}^2},
\end{eqnarray}
in the confining low momentum region of QCD. From (14) it follows that $M(0)\ =\
\frac{gm}{\sqrt{2}\pi}$ and this shows that chiral symmetry is broken in the confining
region of QCD described by the model of Ref.6. The result consists of two parameters,
the dual gluon mass $m$ and the QCD coupling $g$ in the infrared region. We shall
denote $\frac{m^2g^2}{2{\pi}^2}$ by $D$ and estimate its value by evaluating the quark
condensate. The relation between the light quark condensate $<\bar{q}q>$ and $M(p^2)$ for
three colors [26,27] is 
\begin{eqnarray}
<\bar{q}q>&=&-\frac{12}{16{\pi}^2}\int dp^2 \frac{p^2M(p^2)}{A^2(p^2)p^2+M^2(p^2)}.  
\end{eqnarray}
Using (14), we have
\begin{eqnarray}
<\bar{q}q>&=&-\frac{12}{16{\pi}^2D}\int xdx\sqrt{(D-4x/9)}, \nonumber \\
          &=&-\frac{81D\sqrt{D}}{80{\pi}^2}.
\end{eqnarray}
Using the lattice QCD estimate [28] of $<\bar{q}q>=-(250MeV)^3$ for light quarks, 
we find $\sqrt{D}=534 MeV$, which is
also $M(0)$. 
The numerical value of $M(0)$ is 0.534GeV which compares well with lattice
estimate of $\simeq 0.6GeV$ [29].   

\vspace{0.5cm}

Now we use the above results to predict a value for the dual gluon mass $m$. Since our determination
of $D$ using light quark condensate fixes $\frac{m^2g^2}{2{\pi}^2}$ as $0.28 GeV^2$, we need to know $g^2$
for the determination of $m$. We use the expression of Alkofer and Fischer [21] for $\alpha(p^2)$
which is $g^2/(4\pi)$ after normalizing the parameters to $\alpha(M^2_Z)=0.112$. Clearly then $m$
depends upon $p$. In Table.1 we give the numerical values of $m$.

\vspace{0.5cm}

\begin{center}

\vspace{0.3cm}

Table.1

\vspace{0.5cm}

Dual Gluon Mass Vs $\alpha(p^2)$

\vspace{0.3cm}

\begin{tabular}{|c|c|c|}  \hline 
p(GeV) & $\alpha(p^2)$ & Dual Gluon Mass (GeV) \\ \hline 
       &               &  \\
0      &2.972 & 0.384 \\
0.1    &2.770 & 0.398 \\
0.2    &2.317 & 0.436 \\
0.3    &1.852 & 0.487 \\
0.4    &1.487 & 0.544 \\
0.5 &   1.221 & 0.600 \\
0.6    &1.030 & 0.654 \\
0.7    & 0.889& 0.703 \\
0.8    &0.784 & 0.748 \\ 
0.9    &0.703 & 0.790 \\
1.0    &0.641 & 0.828 \\ \hline  
\end{tabular}
\end{center} 
From the Table, we find for $g^2\simeq 8$, the dual gluon mass is $\simeq 828 MeV$, which agrees well
with the estimate of Baker, Ball and Zachariasen [30] who used the above value of $g^2$.   

\vspace{0.5cm}

To summarize, we have used the $\frac{1}{q^4}$ behaviour of the infrared gluon propagator obtained
from the dual Meissner effect description of QCD in the confining region [6] in the Schwinger-Dyson
equation for the quark propagator in the rainbow approximation in Feynman gauge. The chiral symmetry
breaking is seen through $M(0)\neq 0$. Using the light quark condensate value, the numerical value of
the combination $m^2g^2$ is determined as $\frac{m^2g^2}{2{\pi}^2}=0.28 GeV^2$. We have used the
expression of Alkofer and Fischer [21] to evaluate $g^2$ and this depends upon the momentum $p$. The
correspponding dual gluon mass values are given in Table.1. For $g^2=8$ (the value used in Ref.30),
the dual gluon mass is 828 MeV which agrees with that of Ref.30. One of the importances of the dual
gluon mass is to observe that in the rigid string action for QCD derived in Ref.8 starting from Ref.6,
the strength of the extrinsic curvature term is $\frac{{\Lambda}^2}{18\pi m^2}$ which becomes in view
of the above results $\frac{{\Lambda}^2g^2}{36{\pi}^3(0.28GeV^2)}\ =\
\frac{{\Lambda}^2\alpha}{9{\pi}^2(0.28GeV^2)}$, where ${\Lambda}^{-1}$ is a measure of the thickness
of the QCD-string world sheet.   
Since the strong coupling $\alpha$ is asymptotically
free, the extrinsic curvature coupling will also be asymptotically free and this conclusion agrees
with the results of our earlier work on extrinsic curvature coupling [11].     

\vspace{0.5cm}

By using our infrared gluon propagator, the quark-antiquark
potential is found to be $V(r)=\frac{m^2g^2}{8{\pi}^2}r$ and so the string tension is $\sigma =
\frac{m^2g^2}{8{\pi}^2}$. Using the value of $\frac{m^2g^2}{2{\pi}^2}=0.28GeV^2$, as obtained above,
the string tension is estimated for light quarks as 0.07$GeV^2$ and this agrees with the recent
estimate of Weda and Tjon [31]. In the action for the rigid string for QCD in [8] derived from
the dual Meissner effect description, the extrinsic curvature coupling   
in terms of the dual gluon mass and $\Lambda$ is given in the previous paragraph. Using $\sigma =
0.07 GeV^2$, we find for $p=0.4, 0.6, 0.8\ and\ 1.0\ GeV$, the extrinsic coupling (dimensionless) as
$10.24, 2.33, 0.977\ and\ 0.401$ respectively. These
estimates show the importance of the extrinsic curvature coupling in the low momentum confining
region. 

\vspace{0.5cm}

{\noindent{\bf{Acknowledgements}}}

\vspace{0.5cm}

Useful discussions with Ramesh Anishetty are acknowledged with thanks.

\vspace{0.5cm}

{\noindent{\bf{References}}}

\vspace{0.5cm}

\begin{enumerate}
\item H.Araki, K.Hepp and D.Ruelle, Helv.Phys.Acta {\bf{35}} (1962) 164. 
\item F.Strocchi, Phys.Lett. {\bf{62B}} (1976) 60.
\item G.'t Hooft, Nucl.Phys. {\bf{B190}} (1981) 455.
\item S.Mandelstam, Phys.Rev. {\bf{D19}} (1978) 2391.
\item Y.Nambu, Phys.Rep. {\bf{C23}} (1975) 250.
\item R.Parthasarathy, Mod.Phys.Lett. {\bf{A15}} (2000) 2037.
\item A.Aurilia and Y.Takahashi, Prog.Theor.Phys. {\bf{66}} (1981) 693.
\item R.Parthasarathy, Nucl.Phys. (Proc.Suppl) {\bf{B94}} (2001) 562.
\item A.M.Polyakov, Nucl.Phys. {\bf{B268}} (1986) 406. 
\item H.Kleinert, Phys.Lett. {\bf{B174}} (1986) 335.
\item K.S.Viswanathan, R.Parthasarathy and D.Kay, Ann.Phys.(N.Y) {\bf{206}} (1991) 237; \\
      K.S.Viswanathan and R.Parthasarathy, Phys.Rev. {\bf{D51}} (1995) 5830; \\
      R.Parthasarathy and K.S.Viswanathan, J.Geom. and Phys. {\bf{38}} (2001) 207; \\
      R.Parthasarathy and K.S.Viswanathan, Lett.Math.Phys. {\bf{48}} (1999) 243.   
\item H.Kleinert, Phys.Lett. {\bf{B211}} (1988) 151.
\item H.Kleinert, Phys.Lett. {\bf{B293}} (1992) 168. 
\item A.M.Polyakov, Nucl.Phys. {\bf{B486}} (1997) 23. 
\item V.P.Nair and C.Rosenzweig, Phys.Lett. {\bf{B135}} (1984) 450. 
\item C.D.Roberts and S.M.Schmidt, Prog.Part.Nucl.Phys. {\bf{45}} (2000) S1.
\item V.N.Gribov, {\it{QCD at large and short distances. (annotated version)}}, hep-ph/9807224.
\item J.M.Cornwell, Phys.Rev. {\bf{D26}} (1982) 1453.
\item A.C.Mattingly and P.M.Stevenson, Phys.Rev.Lett. {\bf{69}} (1992) 1320.
\item A.C.Aguilar, A.Mihara and A.A.Natale, Int.J.Mod.Phys. {\bf{A19}} (2004) 249.
\item R.Alkofer and C.S.Fischer, {\it{The Kugo-Ojima confinement criterion and the Infrared Behaviour
of Landau gauge QCD}}, hep-ph/0309089; C.S.Fischer and R.Alkofer, Phys.Rev. {\bf{D67}} (2003) 094020.
\item T.Kugo and I.Ojima, Phys.Lett. {\bf{73B}} (1978) 459; Prog.Theor.Phys. {\bf{60}} (1978) 1869.
\item R.Anishetty and S.K.Kudtarkar, Phys.Lett. {\bf{B569}} (2003) 175. 
\item H.J.Munczek and A.M.Nemirovsky, Phys.Rev. {\bf{D28}} (1983) 181. 
\item P.Maris and C.D.Roberts, Int.J.Mod.Phys. {\bf{E12}} (2003) 297; \\
      A.Krassnig and C.D.Roberts, {\it{Dyson-Schwinger Equations: An instrument for hadron physics}},
nucl-th/0309025. \\
      M.S.Bhagawat, A.Hoell, A.Krassnig, C.D.Roberts and P.C.Tandy, {\it{Aspects and Consequences of a
dressed quark-gluon vertex}}, nucl-th/0403012.
\item I.A.Shushpanov and A.V.Smilga, {\it{Quark condensate in a magnetic field}}, hep-ph/970508.
\item Z.G.Wang and S.L.Van, {\it{Calculation of the quark condensate through Schwinger-Dyson
equation}}, hep-ph/0212329.
\item T.R.Hemmert, M.Procura and W.Weise, {\it{Quark Mass dependence of nucleon properties and
extrapolation from lattice QCD}}, hep-lat/0301005. 
\item P.Maris, A.Raya, C.D.Roberts and S.M.Schmidt, nucl-th/0208071. \\
      P.Bowman, V.M.Heller, D.B.Leinweber and A.G.Williams, hep-lat/0209129. \\
      H.Toki, S.Sasaki and H.Suganuma, in {\it{Confinement 95}}, Eds: H.Toki, Y.Mizuno, H.Suganuma,
T.Suzuki and O.Miyamura, World Scientific, 1995. 
\item M.Baker, J.S.Ball and F.Zachariasen, Phys.Rep. {\bf{73}} (1991) 209.
\item J.Weda and J.A.Tjon, {\it{Effects of perturbative exchanges in a QCD-String model}},
hep-ph/0403177.     
\end{enumerate}  
\end{document}